# TOWARDS EHR INTEROPERABILITY IN TANZANIA HOSPITALS: ISSUES, CHALLENGES AND OPPORTUNITIES

Lawrence Nehemiah

School of Computation Communication Science and Engineering, Nelson Mandela African Institution of Science and Technology, Arusha, Tanzania

## ABSTRACT

*This study aimed at identifying the issue, challenges and opportunities from the health consumers in Tanzania towards interoperability of electronic health records. Reaching that level of seamless data sharing among Hospitals needs the cooperation of all stakeholders especially the health consumer whose data are the ones to be shared. Without their acceptance that means there is nothing to share. Recognizing that we conducted a study in Tanzania to identify the challenges, issues and opportunities towards health information exchange through interoperable EHRs. The study was conducted in three major cities of Tanzania to identify the security, privacy and confidentiality issues of information sharing together with related challenges to data sharing. This was in order to come up with a clear picture of how to implement some EHRs that will be trusted by health consumers. The participants (n=240) were surveyed on computer usage, EHRs knowledge, demographics, security and privacy issues. A total of 200 surveys were completed and returned (83.3% response rate). Among them 67.5% were women, 62.6% had not heard of EHRs, 73% were highly concerned about the privacy and security of their information. 75% believed that introduction of various security mechanisms will make EHRs more secure and thus better. We conducted a number of chi-square tests (p<0.05) and we realized that there was a strong relationship among the variable of age, computer use, EHRs knowledge and the concerns for privacy and security.The study also showed that there was just a small difference of 8.5% between those people who think EHRs are safer than paper records and those who think otherwise. The general observation of the study was that in order to make EHRs successful in our Hospitals then the issue of security, and health consumer involvement were they two key towards the road of successful EHRs in our hospitals practices and that will make consumers more willing to allow their records to be shared among different health organizations. So besides the issues identified, this study helped us to identify the key requirements which will be implemented in our proposed framework.*

## 1. INTRODUCTION

The sharing of electronic health information amongst healthcare providers is projected to provide prospective benefits for the healthcare users in the near future due to improved quality, safety, and efficiency in the whole process of care delivery. The successful process of health information sharing will also offer users with new and effective means for cooperating with their healthcare providers than it is now in the managing of their data. These benefits are in line with the strategic goals stated in a report by the Ministry of Health and Social Welfare (MoHSW) in 2013 that is to "enable the health sector to operate more effectively as a connected system, overcoming fragmentation and duplication of service delivery" and also "make patient care safe and effective by ensuring that the correct information is available in a timely manner, where it is needed and to whom it is needed" [1]. With these goals towards increasing the impact of information technology in health sector, then the use of Electronic Health Records is becoming unavoidable. Studies show that it is only a matter of time before the traditional paper-based record keeping is replaced by the electronic one [2]. The target is to establish a shared patient's medical history from the day they are born and throughout their lifetime [3]. However with these promising benefits there are also come some issues especially from the consumers about the safety of the whole process. There are some concerns about security, privacy and confidentiality of patient data that are the key to sharing healthcare information [4]. Some patients are willing for the sharing of their medical records electronically only when they are assured of the privacy and confidentiality.That is, they will not freely feel like giving out any intimate details about their health unless they are assured that no one else who is not involved in their health issues will get their medical records. This is one of the major setbacks that this study has found towards the sharing of medical records in Tanzania through implementation of interoperable EHRs. Some of them say that they prefer paper based health records since it is hard for someone to breach due to its bulky nature. So it is mainly privacy that the consumers are concerned about when it comes to any record system which is electronic. On the other hand there are patient who were concerned about their privacy but also said the advantages that the EHRs are promising outweigh the risks that they may encounter to privacy while others show no problem at all with privacy issues. There are some other technical challenges towards this goal of health information sharing the major one being lack of standardized electronic health records among different healthcare setups. This is caused by the lack of common infrastructure amongst hospital information technology and EHR systems. Another barrier has been identified as the lack of semantics in the interoperability approaches. Semantic interoperability involves the structure of the exchanged data and the coding system involved for the common interpretation of the transferred data [5]. However this study has shown that development in the semantic web technology has some good and promising benefits in trying to solve the issues concerned with semantic interoperability.

## 2. METHODOLOGY

### 2.1 SURVEY DESIGN

In this study about 240 participants were involved, the sample population was taken from three different hospitals in Tanzania located in three different regions; therefore this sample targeted was mainly from those people who were in the process of receiving healthcare and others were those who just accompanied their loved ones to the hospital. The reason for this is due to the fact that these people who were found at the point of care will be more alarmed and attentive of their medical information than those at home for example.

We obtained the participants through a two steps process. The first one was first to locate where we should go for the survey and to obtain a go ahead form the involved hospitals so as to allow their patients and some staff to be involved in the study. We send some introduction letters which were outlining the whole study aim to each of these hospitals (Arusha Lutheran Medical Center, Bugando hospital, and Muhimbili National Hospital) and asking for a permission to use their centers in the survey process. A total of three letters were sent out in which the time required for the whole process to be completed was mentioned. The second step involved preparing the survey itself which involved some questionnaires to be filled out by the participants. We distributed a minimum of 80 questionnaires in each hospital involved. The study took about a month to complete. The response were positive from all the three hospitals involved and we managed to get back a total of 200 responses.

Besides the survey which was mainly done to the health consumers, we were also conducting a simultaneous system study to address some technical issues involved with the systems in the selected hospitals. We did this by observing and oral interview with the systems administrators.

**2.2 DATA ANALYSIS**

In the study we included some demographic questions such as age and sex so as to determine whether there was any relationship that would be clearly attributed to the qualities of such kind of information.

We first analyzed our data by using the standardized descriptive statistics. Later the further analysis was done to point out any statistically significant results by using chi-square analysis. In this analysis the dependent variables in each case were the concern of the participants on health information sharing through EHRs, privacy and security of their medical information and the answers to the question whether they believed that EHRs were more secure and efficient than the paper-based records which is dominant as of now. By using SPSS v20 all the analyses were computed and a 2-tailed significance at the level of $P<0.05$ was considered.

## 3. RESULTS

We were able to get 200 responses which is a response rate of 83.3 %. The demographics outcomeshave revealed that the majority of the people who partook this survey were between the ages of 20 and 40 which indicate that those attending healthcare providers were in their 20s or approaching middle age. There were more women than men and this is just to keep on proving the fact most of those attending the hospitals are women just as some other studies which were done before like the one which was done in New Zealand by Chhanabhai and Holt [6]. In our study there were 135 women which make up 67.5% of all respondents. Most people who participated in this study were regular computer users (80%), e-mail users were (78%), and internet users were (72%).

One of the questions that the participants were asked is about their concerns in regard to the confidentiality and privacy of their health records and it was found out that in general the participants were concerned about their health information. The results are as follows:

Table 1: Participants concerns about the confidentiality and privacy of their health records

|  | YES | NO | NOT SURE | TOTAL |
|---|---|---|---|---|
| No. of participants | 146 | 42 | 12 | 200 |
| Percentage | 73 | 21 | 6 | 100 |

We also compiled a list of all likely possible problems that may be faced with the use of EHRs and presented to the participants. The results indicated that the majority of the participants perceived that the EHR will surely lead to easy sharing of their medical records. Though 40% believed that EHRs will also lead to simpledisclosure of their delicate medical information and shared without their consent (39%). The participants were asked of whether the EHRs will increase the occurrence of medical errors and 37% agreed while those who did not agree were 39%. This show that the issue of medicinal errors when using interoperable EHRs was not of so much concern to the health consumers. Participants believed that EHRs will have issues on the access methods with 53% believing that EHRs have an in-built weakness in its security system and 30% believing that EHRs would have a strong security mechanism.

Since the EHRs are implemented in the same platforms as other computer systems we then introduced a list of common problems that are occurring in everyday computer systems and asked the participants to rank them as either they believe or otherwise that those common problems will also be faced when using EHRs. Some of those problems are the manufacturer's access, hackers, viruses and malicious software. We grouped them into three that is agree, neutral and disagree. Table 2 below gives then summary of the results:

Table 2: Participants' perceptions on common problems of computer systems that can also affect EHRs

|  | Agree | Neutral | Disagree | Total (%) |
|---|---|---|---|---|
| Hackers | 79.5 | 5 | 15.5 | 100 |
| Malicious software | 69 | 12 | 19 | 100 |
| Manufacturer access | 70 | 13 | 17 | 100 |
| Backup failure | 51 | 11 | 38 | 100 |

On the introduction of different kinds of security measures that could be implemented with the EHRs, most of the participants agreed that will make the EHRs more secure and more acceptable. Some mechanisms that were introduced to the participants were the use of antiviruses, implementing restricted system access, the use of firewalls, and encryption mechanisms. 75% of the respondents agreed that this will make EHRs more secure and acceptable.

Table 3: The perceptions on whether security will increase on the introduction of different security mechanisms

|  | **Strongly agree** | **Agree** | **Neutral** | **Disagree** | **Strongly disagree** |
|---|---|---|---|---|---|
| Antivirus | 49 | 30.3 | 10.7 | 6.6 | 3.4 |
| Firewall | 60.1 | 28.3 | 5 | 4.3 | 2.3 |
| Restricted access system | 56.7 | 31 | 5 | 4 | 3.3 |
| Audit log | 50.3 | 36 | 7 | 3.7 | 3 |
| Encryption | 50 | 30.4 | 11.3 | 5 | 3.3 |

Due to the fact that most hospitals in Tanzania are still far behind on using IT in their day to day activities and they are still keeping records on papers, we included another quantitative measure as to whether they believed EHRs were more secure than paper records or otherwise. This helped us to see the real reason behind low adoption of EHRs as well as the user acceptance level of the EHR. Due to the nature of our healthcare environment it was assumed that the participants would believe that paper records are still safer and we expected a big difference between these two answers. Surprisingly the results from this study indicated otherwise. There was only a difference of 8.5% between those who favored the paper records and those who favored the EHRs. 53.5% of the contributors believed more in the paper records and among the reasons which were frequentlystatedis that most don't trust computers because they have a habit of crushing while others said they were afraid of hackers. In the general commentaries, hackers were statedaround105 times which makes up 61% of those who believed the paper records were safer. So this brings us back to the same conclusion that most of the participants still believe that the security is not strong enough with EHRs. 45% agreed that the EHRs are safer than paper records and most of them said it was because papers get lost or damaged and sometimes misplaced; this is the main reason mentioned to move to the EHRs.

We carried out a chi-square test on all the questions involved and the dependent variables were participant's concern for privacy and security of their records, participants' answers to the questions of paper records safety against electronic ones.

The following is what was found out from the test where at each case $P<0.05$:

  i. Most of the younger age group didn't care so much about the security concerns due to the fact that they were more familiar with computers ($P=0.02$)
 ii. Worries of security was mainly shown by women group since they are the greater users of the healthcare ($P=0.06$)
iii. Consumers agreed that EHRs will provide most of the benefits that have been proposed since they are consumer-focused ($P=0.04$).
iv. Consumers when introduced to a list of negative aspects of EHRs, will agree that EHR may rise a number of concerns for them ($P=0.03$).
 v. When consumers are made aware of the security measures that may be in place, they believe that their records will be more secure ($P=0.01$).

## 4. DISCUSSION AND CONCLUSION

From our study we observed that the issue of security and privacy have been then major concern of the health consumers, and this issue is not only for Tanzania but worldwide. According to Dechene, he conducted a study on the barriers to health information sharing and found out that most health consumers will not accept the idea of sharing their medical records if there will not be proper security and privacy mechanisms in place [7]. These two concerns are mentioned as the most contributors to the limited growth of EHRs and their acceptance by consumers. This is very important to address because it is one of the barriers that keeps most if the hospitals from sharing their medical records. There were some previous studies like this one which came up with the same observation [8]. Those studies did show that EHRs pose a problem when it comes to keeping their information private and confidential. From our study similar opinions can be observed. Not only that the health consumers are concerned about privacy and security but also about unauthorized access by some people. A recent study by Butler concluded that most patient will be unwilling to share their medical information if trust is lacking and they believed that the most part of the trust should come from the healthcare providers themselves [9]. Therefore we can conclude that the Tanzania health consumers should be made comfortable by ensuring that the issue surrounding privacy and security of their health records are clearly addressed before taking any further step towards the implementation of interoperable EHRs for health information exchange. This was also observed by a study by Dimitropoulos where he concluded that in order to ensure a wide spread of health consumer participation in the whole process of health information sharing then their participation in determining how the process should take placeis very crucial regardless of their varying levels of privacy and security concerns [10].

We have also showed that many of the participants in the study were worried about hackers who can get access to their health information, but the reality is only 20% of the unauthorized data access are coming from hackers [11]. Most of the access breaches are in fact the job of the inside personnel or the insufficiency of the operational policies in most of the hospitals. Just as a simple example in one of the hospitals that I went for a field study I was allowed to access the real patient data with names and some details. So in this regards we propose that it is important to educate the consumers as to where the impending dangers come from and how they can be dismissed and monitored.

Furthermore we have shown that there was a slightly small difference in the total outcome between the participants who believed paper records were safer than the EHRs and those who believed otherwise. From what most of them said it can be concluded that the only thing that will increase more trust to the EHRs will be the improved security mechanisms. This was a bit surprising result since it was hypothesized that the difference between the two answers would have been bigger. That is due to the nature of our healthcare environment we expected most people would say paper records are safer than EHRs and the difference would have been so big but the difference was just 8.5%. This may lead us to assume that if the health consumers are aware of the security mechanisms and they are a little familiar on how they are operating then their perceptions of security of EHRs will be increased and they will tend to trust more on the electronic records.

We also found out that most participants believed that all the advantages promised by the EHR would be beneficial to the health system. This was supported by a number of replies in the comments where some of them said the advantage of EHRs far outweigh their disadvantages and also some others said if they could be assured of the security measures which are correctly implemented then they will then turn to the EHRs. So this indicates that not only do the consumer know about the advantages that are found in EHRs and can never be found in the paper records but also they are willing to share their records electronically if they are assured of the security to their information. The same results were found in a similar study which was done in New Zealand in 2007 by Chhanabhai et al [12]. In their study about 60% believed paper records were safer than electronic ones but they said if the EHRs are given strong security mechanisms then they will consider EHRs safer than paper records.

In any field involving technology user acceptance is very important even than the technology itself and that is even more important in health sector and these results have shown us that it is not that the EHRs are not accepted but there are some areas especially security wise that needs to be addressed first. So there is a need of proper measures to be put in place to ease the worries of consumers and increase the acceptance level of the EHRs. Sometimes the worries are caused simply by the traditionaldissimilarities, newness to the technology, or merely the sense that health information is too personal to be shared among different healthcare providers.

## 5. THE STUDY LIMITATIONS

Time of the study was crucial. This study was limited to less than a year and that is why we only included the participants from only three major cities and we didn't go up to the rural areas. So for any future study like this it is advised to have more time to reach even the villages and maybe some location factors and a larger population sample may play a significant role in the results. However even though the time was very limited we were able to achieve what we aimed for the study.

## 6. ACKNOWLEDGEMENTS

We would like to thank all the people and the healthcare organizations who participated in this study.

International Journal of Computer Science, Engineering and Applications (IJCSEA) Vol.4, No.4, August 2014## REFERENCES

[1] MoHSW. (2013). Tanzania National eHealth Strategy June, 2013 – July, 201 8. Available: www.tzdpg.or.tz/fileadmin/documents/dpg_internal/dpg_working_groups_clusters/cluster_2/health/Key_Sector_Documents/Tanzania_Key_Health_Documents/Tz_eHealth_Strategy_Final.pdf. Last accessed 31st July 2014.

[2] Gillies J, Holt A. Anxious about electronic health records? No need to be. N Z Med J. 2003; 116:U604. [PubMed: 14581956]

[3] Gordon, A. (2012). Accelerating Electronic Information Sharing to Improve Quality and Reduce Costs in Health Care. Available: http://www.neilstoolbox.com/bibliography-creator/reference-website.htm. Last accessed 26th July 2014.

[4] Francis, L. (2012). When patients interact with EHRs: problems of privacy and confidentiality. Houston Journal of Health Law & Policy. 12 (n), p171-199.

[5] Roney, K. (2012). If Interoperability is the Future of Healthcare, What's the Delay? Available: http://www.beckershospitalreview.com/healthcare-information-technology/if-interoperability-is-the-future-of-healthcare-whats-the-delay.html. Last accessed 20th July 2014.

[6] Chhanabhai, P., Holt, A. (2007). Consumers Are Ready to Accept the Transition to Online and Electronic Records If They Can Be Assured of the Security Measures. Medscape General Medicine. 9 (1), p1-8.

[7] Dechene, J. (2010). The Challenge of Implementing Interoperable Electronic Medical Records. Annals of Health Law. 19 (1), p195-204.

[8] Francis, L. (2012). When patients interact with EHRs: problems of privacy and confidentiality. Houston Journal of Health Law & Policy. 12 (n), p171-199.

[9] Butler, M. (2014). Consumers Might Withhold Information if Trust in EHRs is lacking. Available: http://journal.ahima.org/2014/03/21/consumers-might-withhold-information-if-trust-in-ehrs-is-lacking/. Last accessed 24 July 2014.

[10] Dimitropoulos, L., Patel, V., MApSt, S., Posnack, S. (2011). Public Attitudes toward Health Information Exchange: Perceived Benefits and Concerns. THE AMERICAN JOURNAL OF MANAGED CARE. 17 (n), p111-116.

[11] Chhanabhai, P., Holt, A. (2007). Consumers Are Ready to Accept the Transition to Online and Electronic Records If They Can Be Assured of the Security Measures. Medscape General Medicine. 9 (1), p1-8.

[12] Chhanabhai, P., Holt, A. (2007). Consumers Are Ready to Accept the Transition to Online and Electronic Records If They Can Be Assured of the Security Measures. Medscape General Medicine. 9 (1), p1-8.
36